\pdfoutput=1
\documentclass[aps,prl, twocolumn, showpacs,amsfonts,amsmath,floatfix, superscriptaddress]{revtex4-1}
\usepackage{amsmath}
\usepackage{amssymb}
\usepackage{dsfont}
\usepackage{bm}
\usepackage{graphicx}
\usepackage{pict2e}
\usepackage{xspace}
\usepackage{titlesec}
\usepackage{hyperref}
\usepackage[dvipsnames]{xcolor}
\usepackage{physics}
\usepackage{braket}
\usepackage{siunitx}
\usepackage{float}

\graphicspath{{./figures/}}

\newcommand{\prlsection}[1]{\textit{#1.---}\hspace{+0em}}
\newcommand{\bc}{\begin{center}}
\newcommand{\ec}{\end{center}}
\newcommand{\beq}{\begin{equation}}
\newcommand{\eeq}{\end{equation}}
\newcommand{\beqq}{\begin{equation*}}
\newcommand{\eeqq}{\end{equation*}}
\newcommand{\beqa}{\begin{align}}
\newcommand{\eeqa}{\end{align}}
\newcommand{\barr}{\begin{array}}
\newcommand{\earr}{\end{array}}
\newcommand{\bi}{\begin{itemize}}
\newcommand{\ei}{\end{itemize}}

\newcommand{\s}{\ensuremath{\sigma}}

\newcommand{\cc}[1]{\overline{#1}}
\begin{document}

\title{Universal Gate Set for Continuous-Variable Quantum Computation\\with Microwave Circuits}

\author{Timo Hillmann}
\affiliation{Department of Microtechnology and Nanoscience (MC2), Chalmers University of Technology, SE-412 96 Gothenburg, Sweden}
\affiliation{Institut f\"ur Theorie der Statistischen Physik, RWTH Aachen, 52056 Aachen, Germany}
\author{Fernando Quijandr\'{\i}a}
\affiliation{Department of Microtechnology and Nanoscience (MC2), Chalmers University of Technology, SE-412 96 Gothenburg, Sweden}
\author{G\"oran Johansson}
\affiliation{Department of Microtechnology and Nanoscience (MC2), Chalmers University of Technology, SE-412 96 Gothenburg, Sweden}
\author{Alessandro Ferraro}
\affiliation{Centre for Theoretical Atomic, Molecular and Optical Physics, Queen's University Belfast, Belfast BT7 1NN, United Kingdom}
\author{Simone Gasparinetti}
\affiliation{Department of Microtechnology and Nanoscience (MC2), Chalmers University of Technology, SE-412 96 Gothenburg, Sweden}
\author{Giulia Ferrini}
\affiliation{Department of Microtechnology and Nanoscience (MC2), Chalmers University of Technology, SE-412 96 Gothenburg, Sweden}

\date{\today}

\begin{abstract}
We provide an explicit construction of a universal gate set for continuous-variable quantum computation with microwave circuits. 
Such a universal set has been first proposed in quantum-optical setups, but its experimental implementation has remained elusive in that domain due to the difficulties in engineering strong nonlinearities.
Here, we show that a realistic three-wave mixing microwave architecture
based on the  SNAIL [Frattini \textit{et al.}, Appl. Phys. Lett. \textbf{110}, 222603 (2017)]
allows us to overcome this difficulty.
As an application, we show that this architecture allows for the generation of a cubic phase state with an experimentally feasible procedure. This work highlights a practical advantage of microwave circuits with respect to optical systems for the purpose of engineering non-Gaussian states, and opens the quest for continuous-variable algorithms based on few repetitions of elementary gates from the continuous-variable universal set.
\end{abstract}

\maketitle 
\prlsection{Introduction}
\label{sec-intro} 
The ability to control and manipulate quantum systems has reached an unprecedented level in the past decades 
~\cite{haroche2013nobel,wineland2013nobel}. Quantum computation stems as one of the most promising potential applications of this enhanced controllability of quantum systems~\cite{wendin2017quantum,haeffner_quantum_2008, Arute:2019aa}.
As an alternative to the use of two-level systems for quantum information encoding, continuous-variable (CV) architectures have emerged, where the underlying hardware consists in quantized radiation, either with optical devices~\cite{haroche2013nobel}, microwaves~\cite{ofek2016extending}, or in cavity optomechanics~\cite{Aspelmeyer-RevModPhys}.

The theoretical setting for quantum computation with CV-based architectures has been laid down  in a seminal paper by Lloyd and Braunstein~\cite{Lloyd1999}. There,  universal quantum computation in CVs is defined as the ability of implementing any evolution corresponding to Hamiltonians that are arbitrary polynomials in the mode quadratures. %
The basic ingredients to achieve CV universality are a set of Gaussian gates and  a single non-Gaussian gate, which can be chosen arbitrarily among the polynomials of degree higher than 2 in the quadratures of the quantized modes. The ability to perform arbitrary sequences of these elementary quantum gates ensures universal quantum computation~\cite{Lloyd1999}. 

Since then, the community of quantum opticians has devoted considerable theoretical as well as experimental effort towards developing the building blocks for CV universality. 
In this framework, the experimental challenge consists in achieving a non-Gaussian operation. Experimental effort has focused on photon subtraction~\cite{parigi2007probing,ourjoumtsev2006generating,zavatta2008subtracting,Ra2017,ra2019non,deJ.Leon-Montiel:18, Marek2011, Yukawa2013}, 
photon detection~\cite{eaton2019non,yukawa2013generating, lamperti2014generation} and on the use of ancillary low photon-number states combined with homodyne detection~\cite{etesse2015experimental, ourjoumtsev2007generation} as ways to achieve probabilistic non-Gaussian transformations, resulting however in low success probabilities~\cite{Arzani2017} and limited versatility. In particular, much of the effort has been dedicated to engineer the so-called ``cubic phase gate", 
or alternatively to generate a ``cubic phase state" ~\cite{gottesman_encoding_2001}.  Availability of the latter state allows for engineering a cubic phase gate by gate teleportation~\cite{gottesman_encoding_2001,Marek2011, Yukawa2013, Miyata2016, Arzani2017,Yanagimoto2019,weedbrook2012gaussian}, and thereby promotes the set of Gaussian operations to a universal set~\cite{Lloyd1999, gu2009quantum}. Having at disposal such a cubic gate would allow in particular to engineer Gottesman-Kitaev-Preskill (GKP) states~\cite{gottesman_encoding_2001,Douce2019}, which have been shown to yield fault-tolerance~\cite{menicucci2014fault, gottesman_encoding_2001, fukui2018high, vuillot_quantum_2019, noh_fault-tolerant_2020}. 
Despite these efforts, the generation of a cubic phase state, as well as the implementation of a cubic phase gate, have remained elusive, due to the weakness of the nonlinearities that are available in the optical regime.
Alternatively, deterministic nonlinear gates in strongly coupled quantum electrodynamics
(QED) setups~\cite{park2018deterministic} as well as 
the dissipative stabilization of cubic phase states in optomechanical systems have been proposed~\cite{Houhou2018,Brunelli2019}.

In microwave quantum optics, commonly referred to as circuit QED (cQED), 
nonlinear photon interactions are made possible via the use of Josephson
junctions-based devices~\cite{Gu2017}.
Through a high degree of control in their interactions with linear resonators, they
led to prominent realizations such as the generation of arbitrary Fock states 
and their superpositions~\cite{Hofheinz2009}, Schr\"odinger cat states~\cite{Vlastakis2013, lescanne_exponential_2020} and Gottesman-Kitaev-Preskill states~\cite{campagne2019stabilized}.

In particular, the Superconducting Nonlinear Asymmetric Inductive eLement (SNAIL)~\cite{frattini_3-wave_2017,frattini_optimizing_2018, sivak_kerr-free_2019} enables the mediation of degenerate and non-degenerate three-photon interactions without spurious effects commonplace in cQED through a combination of its geometry and control through external magnetic fields.

The CV notion of universality has not been studied thoroughly in these systems yet, nor has it ever been achieved experimentally. 
Here, we bridge between the optical and microwave approaches, and show explicitly that a parametrically modulated microwave architecture making use of a SNAIL allows for  implementing a universal gate set for CV quantum computation, in the sense of the CV universality notion recalled above~\cite{Lloyd1999}. As an application, we show that a state-of-the-art microwave platform allows for the generation of a cubic phase state - a long sought-after aspiration for the quantum optics community. 

\prlsection{Universal gate set in continuous-variables}
\label{sec:universal-gate-set} 
A universal gate set for CV quantum computation is provided by the following operations~\cite{Lloyd1999}:
\begin{equation}
\label{eq:universal-gate-set}
\hspace{-6pt} \{ \mathrm{e}^{i \hat q_k s_1}, \mathrm{e}^{i (\hat{q}_k \hat{p}_k + \hat{p}_k \hat{q}_k) s_2}, \mathrm{e}^{i (\hat{p}_k \hat{q}_l - \hat{q}_k \hat{p}_l)}, \mathrm{e}^{i \frac{\pi}{4} (\hat{q}_k^2 + \hat{p}_k^2)}, \mathrm{e}^{i \hat q_k^3 \gamma} \},
\end{equation}
where $\hat q_k =  (\hat a_k +  \hat a^\dagger_k)/\sqrt{2}$ and $\hat p_k =  (\hat a_k -  \hat a^\dagger_k)/(i\sqrt{2})$ are the quadrature operators for mode $k$
satisfying the canonical commutation relation $[\hat q_k, \hat p_l] = i \delta_{kl}$ (from here on $\hbar = 1$ and we drop the mode index if only a single mode is relevant).

The operations in Eq.~\eqref{eq:universal-gate-set} excluding $e^{i \hat{q}^3 \gamma}$ represent respectively the displacement, squeezing, beam-splitter and Fourier-transform operators (where $s_i \in \mathbb R$ for all $i$), and are universal for Gaussian operations, i.e., they allow implementing any arbitrary quadratic Hamiltonian. 
Addition of the cubic phase gate $e^{i \hat{q}^3 \gamma}$, where one value of the cubicity $\gamma$ is sufficient~\cite{Lloyd1999, weedbrook2012gaussian, Douce2019, albarelli_resource_2018, takagi2018convex, Supp-Inf}, allows promoting the Gaussian set of operations to universal quantum computation. Indeed, following~\cite{Lloyd1999}, if we can apply the Hamiltonians $\hat A$ and $\hat B$ for a time $\delta t$, then we can approximate the evolution under their commutator for a time $\delta t^2$ by means of the relation
\begin{equation}
\label{eq:comm-approx}
{\rm e}^{-[\hat A, \hat B] \delta t^2} = {\rm e}^{i \hat A \delta t}  {\rm e}^{i \hat B \delta t}  {\rm e}^{-i \hat A \delta t}  {\rm e}^{-i \hat B \delta t} + \mathcal{O}(\delta t^3).
\end{equation}
Commuting a polynomial in $\hat q$ and $\hat p$ with $\hat q$ and $\hat p$ themselves reduces the order of the polynomial by at least 1. Commuting with quadratic Hamiltonians never increases the order.  Finally, commuting with a polynomial of order 3 or higher increases the order by at least 1. Hence, one can achieve arbitrary Hermitian polynomials of any order in $\hat q$ and $\hat p$ by commuting properly the Gaussian operations with an applied Hamiltonian of order 3 or higher. Consequently, the set in Eq.~\eqref{eq:universal-gate-set} is universal. Direct application of Eq.~\eqref{eq:comm-approx} may result in a significant number of operations in order to approximate a desired Hamiltonian. More efficient schemes involving nested operations as well as numerical optimization may provide shorter gate sequences for achieving the same approximate Hamiltonian evolution~\cite{Sefi2011,Sefi2013, arrazola_machine_2019}. However, the approach described above will suffice for our purpose of establishing a proof of principle for universality with microwave circuits.
We are now going to introduce a cQED architecture that is instrumental to
implement the universal gate set in Eq.~\eqref{eq:universal-gate-set}.

%
\prlsection{Microwave circuit for CV universal quantum computation}
\label{sec:from-model-to-gates}
Interactions between microwave photons in superconducting circuits can be realized by coupling the modes of interest to Josephson junctions acting as non-linear, low-loss inductive elements with potential energy $U(\varphi)=E_J(1- \cos(\varphi))$, where $\varphi$ is the superconducting phase across the junction 
and $E_J$ is the Josephson energy~\cite{Devoret2004, Gu2017}. 
When Josephson junctions are arranged in a loop configuration, as in a dc superconducting quantum interference device (SQUID), the Josephson potential energy depends on the magnetic flux threading the loop, allowing for in-situ static tuning of the potential, as well as its parametric modulation~\cite{Wustmann2013,Wustmann2017}. Photon-photon interactions have been demonstrated in resonators terminated by dc-SQUIDs by introducing suitable drives to resonantly select specific $n$-photon processes from the Josephson potential. This potential has even parity for both a single junction and a symmetric SQUID, resulting in mixing processes with an even number of photons, such as four-wave mixing~\cite{Svensson2017}. 
As it has been recently demonstrated, 
an asymmetry between the SQUID junctions introduces an odd contribution to the potential, enabling three-wave mixing (as well as higher-order odd photon interactions)~\cite{Svensson2018, Sandbo2019}. 
However, the even contribution still results in undesired terms, most notably, self- and cross-Kerr interactions, that contain an equal number of creation and annihilation operators and are consequently resonant (non-rotating) in any reference frame. 
To overcome this challenge, the SNAIL 
was recently introduced~\cite{frattini_3-wave_2017,frattini_optimizing_2018, sivak_kerr-free_2019} in the context of Kerr-free three-wave mixing and parametric amplification. 
Here we propose a tunable resonator design based on a SNAIL, and show that by a two-tone flux modulation we can resonantly select all processes comprising the cubic interaction $(\hat a + \hat a^\dagger)^3$ as we will detail later.
\begin{figure}[!bp]
	\centering
	\includegraphics[scale=1]{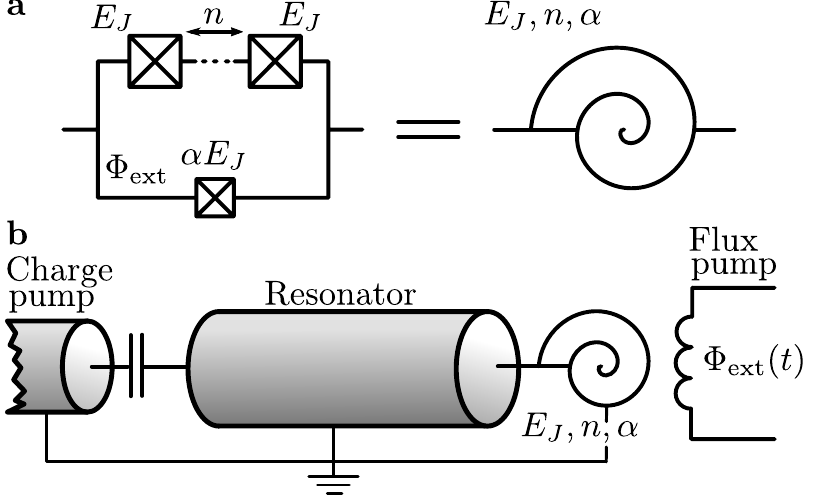}
	\caption{\textbf{a} Circuit representation of the SNAIL composed of $n$ large Josephson junctions of energy $E_J$ and a single smaller one of energy $\alpha E_J$.
	Following~\cite{frattini_3-wave_2017} we represent this subcircuit by a snail-like symbol. \textbf{b} Sketch of our proposed architecture. The quarter wavelength transmission line resonator is terminated into a SNAIL at the right end and capacitively coupled to an input transmission line at the left end
	through which microwave signals for control can be fed.
	The SNAIL potential can be tuned and modulated through an external flux line. 
	}
	\label{fig:snail-circuit}
\end{figure}
The SNAIL loop consists of $n$ large Josephson junctions in parallel with
a single smaller junction with Josephson energies $E_J$ and $\alpha E_J$ ($\alpha < 1$), respectively 
(Fig.~\ref{fig:snail-circuit}a). By threading an external magnetic flux $\Phi_{\rm ext}$ through the loop, the inductive energy of the SNAIL circuit can be written as~\cite{frattini_3-wave_2017}
\begin{align}
\label{eq:SNAIL-potential}
U_{\mathrm{SNAIL}}(\varphi) = - \alpha E_J \cos(\varphi) - n E_J \cos( \frac{\varphi_{\mathrm{ext}} - \varphi}{n}),
\end{align}
where $\varphi$ is the superconducting phase across the small junction, $\varphi_{\mathrm{ext}} = 2\pi \Phi_{\rm ext}/ \Phi_0$ is the reduced applied magnetic flux
and $\Phi_0$ is the magnetic flux quantum. 
The advantage of the SNAIL circuit over traditional SQUIDs is that through its design, resulting in specific parameters $n$, $\alpha$, in addition to  
the external flux $\Phi_{\mathrm{ext}}$, 
its potential landscape around a minimum $\varphi_{\min}$ can be tailored.
Then we can Taylor expand~\eqref{eq:SNAIL-potential}
around this value as $U_{\mathrm{SNAIL}}(\varphi) = U_{\mathrm{SNAIL}}(\varphi_{\rm min}) + \sum_{m>1} c_m (\varphi - \varphi_{\rm min})^m/m!$. The three-wave mixing capability of this device corresponds to setting the coefficient of the fourth-order term identical to zero ($c_4 = 0$)~\cite{frattini_3-wave_2017}, so that the leading 
nonlinear term
is the cubic one.

Our proposed architecture is a quarter wavelength transmission line resonator of length $d$ terminated with a SNAIL loop in one end (Fig.~\ref{fig:snail-circuit}b).
We describe the state of the resonator in terms of the superconducting phase field $\varphi(x,t)$.
For our purposes, the phase at the position of the SNAIL $\varphi_S = \varphi(d,t)$ is assumed to be small. This means that the current flowing through the Josephson junctions is smaller than their corresponding critical currents. 
If this holds, the Lagrange equations of motion which determine the normal modes of the resonator - SNAIL system can be linearized, allowing one to obtain the system Hamiltonian. 
The nonlinear corrections are reintroduced perturbatively~\cite{Supp-Inf}. 
The resonator is also weakly coupled to an input transmission line (Fig.~\ref{fig:snail-circuit}b), which allows 
driving the resonator field.

We choose the SNAIL parameters $n$, $\alpha$ and $\Phi_{\mathrm{ext}}$ in order to operate the device free of Kerr interactions.
Furthermore, we propose to endow this device of flux tunability 
in order to fully exploit the third-order interaction.
For this, we apply a periodically modulated reduced external flux of the form 
\begin{align}
\label{eq:td-flux}
\varphi_{\mathrm{ext}}(t) = \varphi_{\mathrm{ext}}^{\mathrm{dc}} + \varphi_{\mathrm{ext}}^{\mathrm{ac}}(t),
\end{align}
where $\varphi_{\mathrm{ext}}^{\mathrm{dc}}$ is the static part of the flux and $\varphi_{\mathrm{ext}}^{\mathrm{ac}}(t)$ is the time-dependent modulation. 
The latter
satisfies $\lvert  \varphi_{\mathrm{ext}}^{\mathrm{ac}}(t) \rvert \ll 1$. This is required in order to remain near the equilibrium point and to not excite higher nonlinear processes. 
We consider the SNAIL potential up to second order in $\varphi_{\mathrm{ext}}^{\mathrm{ac}}(t)$.
As customary, we follow the canonical quantization recipe 
for the resonator~\cite{Supp-Inf}.
Upon quantization, the Hamiltonian describing our tunable resonator is
\begin{align}\label{eq:snail-full}
    \hat H &= \omega_r \hat a^\dagger \hat a  + g_1(t) \left(\hat{a} +  \hat{a}^{\dagger} \right) + g_2(t) \left(\hat{a} +  \hat{a}^{\dagger} \right)^2  \\\nonumber
    &+  g_3(t) \left(\hat{a} + \hat{a}^{\dagger} \right)^3 + g_4(t) \left(\hat{a} + \hat{a}^{\dagger} \right)^4,
\end{align}
with time-dependent coefficients $g_i(t)$, $i=1,...,4$ to be discussed below.
Here $\omega_r$ is the resonance frequency of the transmission line resonator modified by the presence of the SNAIL.
Due to the modulation of the potential around its minimum, we gain linear and quadratic contributions in addition to the cubic potential. 
Notice that we have also included a quartic contribution. The reason is twofold. First, 
due to the time-dependent modulation the condition $c_4 = 0$ is not always satisfied. 
Second, as demonstrated in~\cite{frattini_optimizing_2018} the cubic interaction 
leads to fourth-order renormalization effects.
In this context, the Kerr-free operation point results from the interplay of third- and fourth-order nonlinearities.
The time-dependence of $g_1$ and $g_3$ is proportional to $\varphi_{\mathrm{ext}}^{\mathrm{ac}}(t)$ 
and that of $g_2$  and $g_4$ to $\varphi_{\mathrm{ext}}^{\mathrm{ac}}(t)^2$.
The perturbative treatment of the nonlinearity leads to the hierarchy $\omega_r \gg \vert g_i \vert$.

\prlsection{Engineering of Gaussian gates}
We start our demonstration of CV universality by showing that this architecture is capable of implementing the Gaussian operations in \eqref{eq:universal-gate-set}.
For this, a modulation of the flux is not required, i.e., $\varphi_{\mathrm{ext}}^{\mathrm{ac}}(t) = 0$ in Eq.~\eqref{eq:td-flux}. In this case, the Hamiltonian~\eqref{eq:snail-full} reduces to $\hat H = \omega_r \hat a^\dagger \hat a +  g_3^{\mathrm{dc}} \left(\hat{a} + \hat{a}^{\dagger} \right)^3 + g_4^{\mathrm{dc}} \left(\hat{a}+ \hat{a}^{\dagger} \right)^4$, with the couplings $g_3^{\mathrm{dc}}, g_4^{\mathrm{dc}}$ depending only on the microscopic parameters of the circuit as well as the static external flux.
In order to engineer a squeezing operation, we apply an off-resonant microwave tone of frequency $\omega_p = 2\omega_r$
through the input transmission line. As discussed in~\cite{sivak_kerr-free_2019}, 
in a frame rotating at the resonator frequency $\omega_r$
the system is described by the effective Hamiltonian $\hat{H}_{\rm sq} = - \frac{i}{2} \left( \xi \hat{a}^{\dagger 2} - \cc{\xi} \hat{a}^2 \right)$
where the parameter $\xi$ depends on 
$g_3^{\mathrm{dc}}$ as well as on the amplitude of the drive.
In particular, choosing $\xi$ real allows us to obtain the squeezing operation in Eq.~\eqref{eq:universal-gate-set}.
This is the basis of SNAIL-based parametric amplification 
~\cite{frattini_optimizing_2018,sivak_kerr-free_2019}.
The Fourier transform $\mathrm{e}^{i \frac{\pi}{4} (\hat{q}^2 + \hat{p}^2)}$ follows from the free evolution of the system, i.e., the evolution under the resonator Hamiltonian $\omega_r \hat a^\dagger \hat a$ in the absence of any external modulation.
As customary, a displacement operation is implemented by means of a microwave tone near resonance with mode $\hat{a}$.
Finally, a tunable beam-splitter interaction
can be achieved by 
coupling two resonator-SNAIL units via a parametrically modulated dc-SQUID or a tunable gap qubit,
or mediating a static nonlinear coupling via time-dependent drivings of both resonators~\cite{Baust2015,Pfaff2017,Collodo2019}.

\prlsection{Engineering a non-linear gate}
The power of our proposal relies on the realization of the interaction term $\hat q^3$ which has been experimentally elusive so far.
In order to engineer such a gate, we exploit the flux tunability of the SNAIL.
In particular, we consider a two-tone modulation of the form
\begin{align}
\label{eq:ac_flux_drive}
\varphi_{\rm ext}^{\rm ac}(t) = \lambda \left[ \cos(\omega_r t) + \cos(3 \omega_r t) \right],
\end{align}
where $\lambda \ll 1$ is a small modulation amplitude.
This is justified by studying the cubic potential in Eq.~\eqref{eq:snail-full} in the 
interaction picture with respect to the free resonator Hamiltonian $\omega_r \hat a^\dagger \hat a$.
Because of the odd parity of the potential, there are no non-rotating contributions. 
The terms that are pure in $\hat{a}^{(\dagger)}$ rotate with
frequency
$\mp 3 \omega_r$ while the mixed terms rotate with $\pm \omega_r$. Thus,
in order to select the full cubic interaction resonantly,
the necessity
to drive with two frequencies $\omega_r$ and $3 \omega_r$
arises.
It must be pointed out that the drive at $\omega_r$ also selects resonantly the linear and the quadratic terms in Eq.~\eqref{eq:snail-full}. However, in the Supplementary Material we show that for a realistic choice of parameters the quadratic term 
is sufficiently suppressed and can thus be neglected. This is not the case for the linear drive. 
However, 
its effect
can be corrected via a displacement of the resonator field and thus we neglect it in the remainder of this letter.

Finally, following the above arguments and in the rotating frame, we isolate the desired cubic interaction
\begin{align}
\label{eq:cpg-Hamiltonian}
\hat{H}_I &= g_3^{\mathrm{ac}} \left(\hat{a} + \hat{a}^{\dagger} \right)^3,	
\end{align}
where the coupling $g_3^{\mathrm{ac}}$ depends on the microscopic parameters of the circuit as well as on the modulation amplitude $\lambda$. 
In principle one could use classical optimization to determine the circuit parameters that result in Hamiltonian couplings $g_i$ which are tailored to specific requirements.
However, here we will restrict ourselves to hand-selected circuit parameters based on state-of-the-art realizations of superconducting circuits to demonstrate our claim.

For the chosen parameters our theory predicts that it is possible to achieve $g_3^{\mathrm{ac}} / 2 \pi \approx \SI{0.3}{\MHz}$ and $\omega_r / 2 \pi \approx \SI{4}{\GHz}$ while the 
Kerr nonlinearity
is tuned to zero~\cite{Supp-Inf}.
This also guarantees that the drive at $3 \omega_r$ 
is sufficiently detuned from the plasma frequency of the Josephson junctions, which are typically on the order of a few ten of $\si{\GHz}$ \cite{masluk_microwave_2012}.
Therefore, 
we have demonstrated that all of the operations in the set \eqref{eq:universal-gate-set}
can be implemented with our proposed architecture.

We emphasize that in principle more gates are directly accessible through our proposal. While this does not matter for the goal of achieving universality, having at disposal customizable high-order gates can lead to substantial advantages when limited to noisy devices with finite coherence time.

\prlsection{Generation of a cubic phase state}
We now address the 
generation of a cubic phase state $\ket{\gamma, r} = \mathrm{e}^{i \gamma \hat{q}^3} \mathrm{e}^{\frac{r}{2} (\hat{a}^{\dagger 2} -  \hat{a}^2)} \ket{0}$ , where $r$ is the real squeezing parameter, $\gamma$ the cubicity of the cubic phase gate applied and $\vert 0 \rangle$ is the photon vacuum state~\cite{weedbrook2012gaussian}.
Due to the weak coupling to the transmission line the main dissipation channel corresponds to internal losses. 
We treat them
within a Gorini-Kossakowski-Sudarshan-Lindblad master equation formalism with jump operator $\hat L = \sqrt{\kappa} \hat{a}$, where $\kappa$ is the single photon loss rate.
Evolving an initial squeezed state  for a time $t_g$ with the Hamiltonian \eqref{eq:cpg-Hamiltonian} results in a cubic phase state with cubicity $\gamma =  g_3^{\mathrm{ac}} \sqrt{8} t_g$, where the factor $\sqrt{8}$ results from the normalization of $\hat q$~\cite{Supp-Inf}.

\begin{figure}[!tb]
	\centering
	\includegraphics[scale=1]{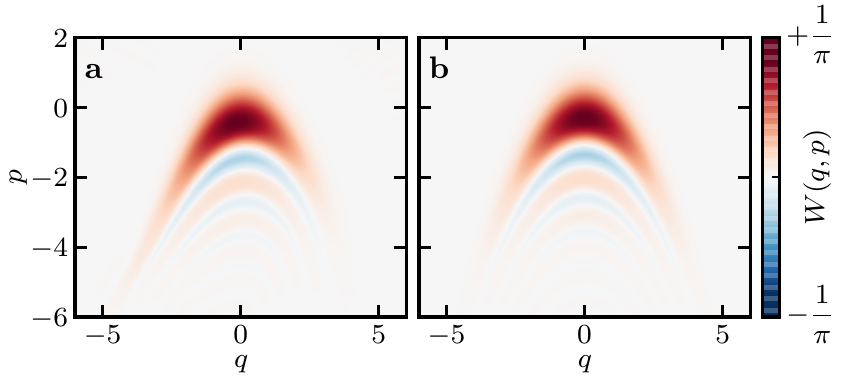}
	\caption{Wigner distribution for the cubic phase state $\ket{\gamma, r}$. \textbf{a}~Obtained from a master equation simulation with the Hamiltonian \eqref{eq:snail-full} and single photon loss rate $\kappa / 2 \pi= \SI{50}{\kHz}$ by sequentially applying the squeezing and cubic phase gate to an initial vacuum state $\ket{0}$ (see~\cite{Supp-Inf} for details). \textbf{b}~Ideal cubic phase state with matched cubicity $\gamma \approx 0.1$ and squeezing $r \approx 0.70 \; (\approx \SI{6}{\dB})$.}
	\label{fig:CPS-Wigner}
\end{figure}

At this point we would like to stress that Hamiltonian~ \eqref{eq:cpg-Hamiltonian} 
is the result of
several approximations, more remarkably, the rotating wave approximation.
In order to test the validity of the latter approximation, we 
can apply the above described squeezing and cubic gates by means of the appropriate external drives to the \emph{full} nonlinear-resonator Hamiltonian Eq.~\eqref{eq:snail-full} initialized in the vacuum state~\cite{Supp-Inf}.
In Fig.~\ref{fig:CPS-Wigner}(a) and (b) we compare the state resulting from the full Hamiltonian~\eqref{eq:snail-full} in the presence of losses with the ideal cubic phase state $\ket{\gamma, r}$, corresponding to the lossless evolution under Hamiltonian~\eqref{eq:cpg-Hamiltonian} acting on an ideal squeezed state, respectively. 
For this example we consider 
$\kappa / 2 \pi = \SI{50}{\kHz}$ ($1/\kappa \approx \SI{3}{\micro\s}$)
which corresponds to a quality factor $Q$ of the order of $10^5$. 
In addition, we
choose evolution times and drive strengths 
that lead to a squeezing parameter $r \approx 0.7$~\footnote{
We would like to emphasize that for any value of squeezing the distinguished features  of the state remain. For example, the negativity of its Wigner function depends on the single effective parameter $\gamma \mathrm{e}^{3r}$~\cite{albarelli_resource_2018}%
}
and cubicity $\gamma \approx 0.1$~\footnote{Having at disposal one specific value of cubicity  is enough for universality. One reason is that arbitrary cubicity value can be achieved complementing the cubic phase gate with additional squeezing gates. In particular, one value of the cubicity that corresponds to implement the $T$-gate within GKP encoding~\cite{gottesman_encoding_2001} is achievable with our method}. 
We obtain a fidelity of \SI{97.2}{\percent} to the ideal cubic phase state
(numerical simulations are done using QuTiP~\cite{johansson_qutip_2013}),
which is sufficient to retain its distinguished features, such as the negativities of the Wigner function, as shown in Fig.~\ref{fig:CPS-Wigner}.
Note that the fidelity in the absence of losses is increased to \SI{97.4}{\percent}.

Finally, the generated cubic phase state can be probed, e.g., by quantum state tomography using a dispersively coupled qubit  (not shown in Fig.~\ref{fig:snail-circuit}(b))~\cite{Hofheinz2009}.

%

\prlsection{Conclusions}\label{secCcl}
In summary, we have proposed a microwave architecture that allows for the implementation of a universal gate set for continuous-variable quantum computation. Our architecture is 
based on a quarter wavelength transmission line resonator terminated by an array of Josephson junctions in a SNAIL configuration. The tunability of our device allows for engineering customized gates, and in particular the interaction $\hat q^3$, corresponding to a cubic phase gate. As an application, we have provided an experimentally realistic protocol for the generation of the cubic phase state, which is a resource state for CV quantum computation, and whose generation has not been experimentally achieved yet despite extensive effort undertaken with quantum optical setups.

On the one hand, our work opens the experimental quest for a cubic phase state with microwave circuits.
Our proposal is within reach of current cQED technology in terms of resonator quality factors, that can be as high as $3 \times 10^5$ in 
3D architectures~\cite{Grimm2019}, and the ability to tune the resonator field much faster than its corresponding lifetime, with pulse synthesis resolution within nanoseconds~\cite{Sandberg2008}.
On the other hand, the experimental realization of a universal gate set in a continuous-variable architecture would present the community with the question what relevant quantum algorithms can be run in the near future on such an architecture, possibly with a limited circuit depth, and without fault tolerance. In this sense, our work heralds further research in an area that could be referred to as of ``Continuous-Variable - Noisy Intermediate Scale Quantum (CV-NISQ) devices", in resonance with similar investigations recently emerged in the context of qubit-based quantum computation~\cite{preskill2018quantum}.
Indeed, quantum advantage in CV beyond a specific encoding has been addressed so far only in the context of sampling problems~\cite{bromley2019applications, Chabaud2017,Douce2017,Douce2019,hamilton2017gaussian,chakhmakhchyan2017boson,lund2017exact}, for the implementation of algorithms for optimization of continuous functions~\cite{verdon2019quantum}, or for numerical integration~\cite{Rebentrost2018}, leaving plenty of room for new applications yet to unveil.
%
\begin{acknowledgments}
We thank Vitaly Shumeiko for useful discussions. G.F. acknowledges support from the VR Grant QuACVA. F. Q., G. J., S. G. and G. F. 
acknowledge the financial support from the Knut and Alice Wallenberg Foundation through the Wallenberg Center for Quantum Technology (WACQT).
T. H. acknowledges support by the
Deutsche Forschungsgemeinschaft via RTG 1995. 
\end{acknowledgments}

%

\end{document}


\title{Supplementary Material to the letter \\``Universal Gate-Set for Continuous-Variable Quantum Computation\\ with Microwave Circuits"}

\author{Timo Hillmann}
\affiliation{Department of Microtechnology and Nanoscience (MC2), Chalmers University of Technology, SE-412 96 Gothenburg, Sweden}
\affiliation{Institut f\"ur Theorie der Statistischen Physik, RWTH Aachen, 52056 Aachen, Germany}
\author{Fernando Quijandría}
\affiliation{Department of Microtechnology and Nanoscience (MC2), Chalmers University of Technology, SE-412 96 Gothenburg, Sweden}
\author{G\"oran Johansson}
\affiliation{Department of Microtechnology and Nanoscience (MC2), Chalmers University of Technology, SE-412 96 Gothenburg, Sweden}
\author{Alessandro Ferraro}
\affiliation{Centre for Theoretical Atomic, Molecular and Optical Physics, Queen's University Belfast, Belfast BT7 1NN, United Kingdom}
\author{Simone Gasparinetti}
\affiliation{Department of Microtechnology and Nanoscience (MC2), Chalmers University of Technology, SE-412 96 Gothenburg, Sweden}
\author{Giulia Ferrini}
\affiliation{Department of Microtechnology and Nanoscience (MC2), Chalmers University of Technology, SE-412 96 Gothenburg, Sweden}

\date{\today}
\maketitle 
\clearpage
\newpage
\setcounter{equation}{0}
\renewcommand{\theequation}{S\arabic{equation}}
\newcommand{\dphi}{\widetilde{\Phi}}
\newcommand{\mphi}{\Phi_{\min}}
\newcommand{\fphi}{\Phi_{\mathrm{ext}}^{\mathrm{ac}}(t)}
\newcommand{\sphi}{\Phi_{\mathrm{ext}}^{\mathrm{dc}}}
\newcommand{\USnail}{U_{\mathrm{SNAIL}}(\Phi)}
\section*{Supplementary Note 1. Circuit Quantization}
\subsection{Inductive Energy of the SNAIL}
The inductive energy of the SNAIL reads
\begin{align}
\label{supp-eq:SNAIL-potential}
U_{\mathrm{SNAIL}}(\Phi) = &- \alpha E_J \cos(\Phi / \phi_0) - n E_J \cos( \frac{\Phi_{\mathrm{ext}} - \Phi}{n \phi_0}),
\end{align}
where $\Phi$ is the flux variable describing the small Josephson junction, $\Phi_{\rm ext}$ is an applied external magnetic flux and
$\phi_0 = \hbar / 2e$ denotes the reduced magnetic flux quantum through which the superconducting phase $\varphi = \Phi / \phi_0$ and flux $\Phi$ are related. Expressing Eq.~\eqref{supp-eq:SNAIL-potential} in terms of the superconducting phase $\varphi$ allows one to obtain Eq.~(3) of the main text.
First of all, we will restrict to 
an external time-independent modulation, i.e., a static (or dc) external flux $\Phi_{\rm ext} = \sphi$.
It is useful to express $\Phi$ as $\Phi = \mphi + \dphi$, where $\mphi$ is the value at which $\USnail$ attains its minimum value. The latter is determined by the condition
\begin{align}
\label{supp-eq:pot-minimum}
\left. \frac{\phi_0}{E_J} \frac{\mathrm{d}}{\mathrm{d} \Phi} \USnail \right\rvert_{\Phi = \mphi} = \alpha \sin(\mphi / \phi_0) - \sin(\frac{\sphi - \mphi}{n \phi_0}) = 0,
\end{align}
which can be solved numerically.
An effective potential in $\dphi$ is obtained by Taylor expanding $\USnail$ around $\mphi$,
\begin{align}
\label{supp-eq:eff-Snail-td}
U_{\mathrm{eff}} (\dphi) / E_J = \sum_{m=2}^{\infty} \frac{c_m^{\rm dc}}{m!} \left(\frac{\dphi}{\phi_0}\right)^m,
\end{align}
where the coefficients $c_m^{\rm dc}$ are time-independent functions of the external parameters $(\alpha, n, \sphi)$ and can be obtained from the $m$th derivative of Eq.~\eqref{supp-eq:SNAIL-potential}, i.e.,
\begin{align}
\label{supp-eq:static-coeffs}
c_m^{\mathrm{dc}} = \left. \frac{\phi_0^m}{E_J} \frac{\mathrm{d}^m}{\mathrm{d} \Phi^m} \USnail \right\rvert_{\Phi = \mphi}.
\end{align}

We now introduce a small time-dependent (or ac) modulation so that the total external flux is
\begin{align}
\label{supp-eq:td-flux}
\Phi_{\mathrm{ext}}(t) = \Phi_{\mathrm{ext}}^{\mathrm{dc}} + \Phi_{\mathrm{ext}}^{\mathrm{ac}}(t),
\end{align}
with $\lvert \Phi_{\mathrm{ext}}^{\mathrm{ac}}(t) / \phi_0 \rvert \ll 1$. This guarantees a perturbative treatment of the potential around the static minimum.
In a similar fashion, the effective potential in the presence of the modulation can be written as
\begin{align}
U_{\mathrm{eff}} (\dphi) / E_J = \sum_{m=1}^{\infty} \frac{c_m(t)}{m!} \left(\frac{\dphi}{\phi_0}\right)^m, \label{supp-eq:eff-Snail} 
\end{align}
with the expansion coefficients $c_m(t)$ becoming time-dependent.
For a small time-dependent modulation, they can be written in the form
$c_m(t) = c_m^{\mathrm{dc}} + c_m^{\mathrm{ac}}(t) $
where the coefficients $c_m^{\mathrm{dc}}$ are determined by Eq.~\eqref{supp-eq:static-coeffs}.
Here we would like to emphasize that $c_1(t) \neq 0$ due to the contribution from $c_1^{\mathrm{ac}}(t) \neq 0$. %

In the presence of the time-dependent modulation,  the full SNAIL potential reads 
\begin{align}
\label{supp-eq:snail-expand}
U_{\mathrm{SNAIL}}(\dphi) / E_J &= - \alpha \left[ \cos(\mphi / \phi_0) \cos(\dphi / \phi_0) - \sin(\mphi / \phi_0) \sin(\dphi / \phi_0) \right] \nonumber \\
& - n \cos(\frac{\sphi - \mphi}{n \phi_0}) \left[ \cos(\frac{ \fphi}{n \phi_0}) \cos(\frac{\dphi}{n \phi_0}) + \sin(\frac{  \fphi}{n \phi_0}) \sin(\frac{\dphi}{n \phi_0}) \right] \nonumber \\
& + n \sin(\frac{\sphi - \mphi}{n \phi_0}) \left[ \sin(\frac{ \fphi}{n \phi_0}) \cos(\frac{\dphi}{n \phi_0}) - \cos(\frac{ \fphi}{n \phi_0}) \sin(\frac{\dphi}{n \phi_0}) \right],
\end{align}
which follows from  trigonometric identities.
From here we derive the time-dependent coefficients $c_m(t)$ by retaining terms up to second-order in $\fphi$.
The first four coefficients that lead to resonant contributions are then given by the relations
\begin{align}
c_1^{\mathrm{ac}}(t) &= -  \cos( \frac{ \sphi - \mphi }{\phi_0 n}) \frac{\fphi}{\phi_0 n}, \qquad
&&c_2^{\mathrm{ac}}(t) = -  \cos( \frac{ \sphi - \mphi }{\phi_0 n}) \frac{\fphi^2}{2! \phi_0^2 n^3},  \label{supp-eq:coeffs-1-2}\\
c_3^{\mathrm{ac}}(t) &=  \cos( \frac{ \sphi - \mphi }{\phi_0 n}) \frac{\fphi}{ \phi_0n^3},  \qquad
&&c_4^{\mathrm{ac}}(t) =  \cos( \frac{ \sphi - \mphi }{\phi_0 n}) \frac{\fphi^2}{2! \phi_0^2 n^5}  \label{supp-eq:coeffs-3-4},
\end{align}
where $c_3^{\mathrm{ac}}(t)= - c_1^{\mathrm{ac}}(t)/n^2$ and $c_4^{\mathrm{ac}}(t) = -c_2^{\mathrm{ac}}(t)/n^2$.
We have explicitly  neglected the quadratic contributions in $\fphi$ for $c_1^{\mathrm{ac}}(t)$ and $c_3^{\mathrm{ac}}(t)$ as they will be off-resonant. 
As discussed in the main text, we choose $\fphi/\phi_0 = \lambda [ \cos(\omega_r t) + \cos(3 \omega_r t)]$ ($\lambda \ll 1$) in order to retain the full cubic potential in an interaction picture (rotating frame). 
As stated above, in the presence of the modulation, the $c_1(t)$ coefficient is not zero. Then, there is a linear contribution $\propto \Phi$ to the effective potential. This is the linear drive discussed in the main text which is resonantly selected by the drive component oscillating at $\omega_r$.

By a similar argument $c_2^{\mathrm{ac}}(t)$ and $c_4^{\mathrm{ac}}(t)$ contain no term that is linear in $\fphi$. 
Indeed, one can show that for a non-zero $c_2(t)$ coefficient, there will be a contribution proportional to $\Phi^2$ selected by the same frequency component. 
Higher resonances selected by the $3 \omega_r$ frequency component will involve higher powers of the flux and, as we discuss in the next section, they can safely be neglected.
In principle, one could choose a sufficiently small value of $\lambda$ in order to get rid of this quadratic effect. Nevertheless, we are dealing with an open system and we require that the effective cubic coupling proportional to $c_3$, and thus proportional to $\lambda$, exceeds the dissipation rate of the system by roughly one order of magnitude.  
%
In the following we will consider only terms with $m \leq 4$ in accordance with recent experiments on SNAIL-circuits~\cite{grimm_kerr-cat_2019, frattini_optimizing_2018, sivak_kerr-free_2019} and similar designs~\cite{lescanne_exponential_2020}.

\newcommand{\flux}{\Phi}
\newcommand{\sflux}{\Phi_S}

\subsection{Transmission Line Resonator and SNAIL Lagrangian}

Here we study a transmission line resonator of length $d$ terminated in an array of $M$ SNAILs. This is a generalization of the setup studied in the main text [Cf. Fig.~{1}b] which corresponds to the case $M=1$.
The Lagrangian of this system is
\begin{align}
\label{supp-eq:system-Lagrangian}
\mathcal{L} &= \frac{1}{2} \int_{0}^{d} \left\{ c \,\left[\dot{\flux}(x, t) \right]^2 - l^{-1} \left[\partial_x \flux(x, t) \right]^2 \right\} \, \mathrm{d}x - M \, U_{\mathrm{eff}} \left(\frac{\sflux}{\phi_0 M} \right),
\end{align} 
with $U_{\mathrm{eff}}$, as specified in Eq.~\eqref{supp-eq:eff-Snail}, which approximates the SNAIL potential around its static minimum. 
The parameters $c$ and $l$ denote the capacitance and inductance per unit length of the resonator (respectively) and the state of the resonator is described by the generalized flux field $\flux (x,t)$.
Finally, $\sflux = \flux (d,t)$ denotes the value of the latter at the SNAIL position. 
We neglect the capacitance of the SNAIL as the charging energy $E_C$ of each junction is negligible compared to the Josephson energy $E_J$ and furthermore assume that the flux drive at $3 \omega_r$ is sufficiently detuned from the plasma frequency of the Josephson junctions.
An extensive discussion of a similar architecture consisting of a half wavelength resonator interrupted by an array of SNAILs at its middle position is given in~\cite{sivak_kerr-free_2019, frattini_optimizing_2018, sivak_josephson_2020}.

We follow now the standard quantization procedure~\cite{sivak_kerr-free_2019,frattini_optimizing_2018, leib_networks_2012, wallquist_selective_2006}. 
To this end we first neglect all nonlinear terms in the Lagrangian and derive the normal mode structure of the system. 
We will then reintroduce the nonlinearities to obtain the Hamilton operator for the system. \\

The linear part of the total Lagrangian reads
\begin{align}
\label{eq:lin-Lagrangian}
\mathcal{L}_{\mathrm{lin}} &= \frac{1}{2} \int_{0}^{d} \left\{ c \,\left[\dot{\flux}(x, t) \right]^2 - l^{-1} \left[\partial_x \flux(x, t) \right]^2 \right\} \, \mathrm{d}x - \frac{E_J}{\phi_0} c_1(t) \sflux -  \frac{E_J c_2(t)}{2 M \phi_0^2} \sflux^2.
\end{align}
The Euler-Lagrange equation
\begin{align}
\label{eq:EL-equation}
\frac{\partial}{\partial t} \frac{\delta \mathcal{L}}{\delta \dot{\flux}} - \frac{\delta \mathcal{L}}{\delta \flux} = 0,
\end{align}
results
in the wave equation 
\begin{align}
\label{eq:wave-equation}
\nu^2 \partial_x^2 \flux (x, t) - \partial_t^2 \flux (x, t) = 0,
\end{align}
for the field in the transmission line resonator,
where $\nu = 1 / \sqrt{c \,l}$ is the phase velocity. 

The structure of the resonator normal modes is determined by the boundary conditions.
We impose that the current at the left end ($x=0$) of the resonator is identical to zero, i.e., 
$-\partial_x \flux (x, t) \rvert_{x=0} / l = 0$. 
On the other hand, the boundary condition at $x = d$ is modified by the presence of the SNAIL array and can be determined by evaluating the Euler-Lagrange equation for $x=d$.
From this we obtain
\begin{align}
\label{eq:boundary-right}
- \partial_x \flux (x, t) \rvert_{x=d} = \frac{l E_J}{\phi_0} c_1(t) +  \frac{l E_J}{M \phi_0^2} c_2(t) \sflux.
\end{align}
We now restrict ourselves to a single mode and make a separation of variables ansatz
\begin{align}
\label{eq:ansatz}
\flux(x, t) = f(x) \phi(t).
\end{align}
Inserting this ansatz into the wave equation \eqref{eq:wave-equation} the problem decouples and we obtain 
\begin{align}
&\partial_x^2 f(x) + k^2 f(x) = 0, \label{eq:ode-space} \\
&\partial_t^2 \phi(t) + \omega_r^2 \phi(t) = 0, \label{eq:ode-time}
\end{align}
with the linear dispersion relation $k \nu = \omega_r$. 
The boundary condition at $x=0$ is satisfied by choosing $f(x) = \cos(k x)$. 
For small modulation amplitudes $\lambda$, it is justifiable to neglect the time-dependence in Eq.~\eqref{eq:boundary-right} altogether such that the normal modes become time-independent. 
Then, inserting $f(x)$ into the boundary condition \eqref{eq:boundary-right} at $x = d$ and dropping any time-dependent terms, we obtain
\begin{align}
\label{supp-eq:eigenmode-eq}
\omega_r \tan(\frac{\pi}{2} \frac{\omega_r}{\omega_0}) = \frac{Z_c c_2}{M L_J},
\end{align}
where 
$\omega_0 = (\pi/2) (\nu / d)$ describes the bare resonance frequency of the resonator in the absence of the SNAIL array and $Z_c = \sqrt{l / c}$, the characteristic impedance of the resonator.
Having obtained a solution for the spatial normal modes we evaluate the integral in Eq.~\eqref{eq:lin-Lagrangian} which simplifies to 
\begin{align}
\label{eq:cos-integral}
\int_0^{d} \cos^2(k x) \, \mathrm{d}x = \frac{d}{2} \left(1 + \frac{\sin( 2 k d)}{2 k d} \right) \equiv \frac{\eta}{c}.
\end{align}
Considering this together with Eq.~\eqref{eq:ode-time}, the linear Lagrangian simplifies to
\begin{align}
\label{eq:lin-Lagragian-simplified}
\mathcal{L}_{lin} = \frac{\eta}{2} \dot{\phi}^2(t) - \frac{1}{2} \eta \omega_r^2 \phi^2(t) - \frac{ E_J c_1(t)}{ \phi_0} \sflux + \frac{E_J c_2^{\mathrm{ac}}(t)}{2 M \phi_0^2} \sflux^2,
\end{align}
with $\sflux = \phi(t) \cos(kd)$. 
\subsection{Nonlinear Terms and the Hamiltonian}
The nonlinear part of the Lagrangian is given by
\begin{align}
\label{eq:Lagrangian-nl}
\mathcal{L}_{nl} = - \frac{E_J c_3(t)}{3! M^2 \phi_0^3} \sflux^3 - \frac{E_J c_4(t)}{4! M^3 \phi_0^4} \sflux^4.
\end{align}
From the total Lagrangian $\mathcal{L} = \mathcal{L}_{\mathrm{lin}} + \mathcal{L}_{\mathrm{nl}}$, we can derive the Hamiltonian from the Legendre transform  and by introducing the conjugate variable 
\begin{align}
N = \frac{\partial \mathcal{L}}{\partial \dot{\phi}} = \eta \dot{\phi}.
\end{align}
As the Legendre transform leaves the nonlinear part of the Lagrangian invariant the Hamiltonian function reads
\begin{align}
\label{eq:Hamiltonian-function}
H(t) = \frac{1}{2 \eta} N^2 + \frac{1}{2} \eta \omega_r^2 \phi^2 + \frac{E_J}{\phi_0} c_1(t) \sflux + \frac{E_J c_2^{\mathrm{ac}}(t)}{2 M \phi_0^2} \sflux^2 + \frac{E_J c_3(t)}{3! M^2 \phi_0^3} \sflux^3 + \frac{E_J c_4(t)}{4! M^3 \phi_0^4} \sflux^4.
\end{align}

The Hamiltonian is quantized by
promoting the fields $\phi$ and $N$ to the operators $\hat \phi$ and $\hat n$ which satisfy the 
commutation relation $[ \hat{\phi}, \hat{N} ] = i \hbar$.
It is convenient to express 
them
in terms of the bosonic annihilation and creation operators $\hat{a}$ and $\hat{a}^{\dagger}$ (respectively) via the relations,
\begin{align}
\label{eq:creation-annihilation}
\hat{N} = - i \sqrt{\frac{ \hbar \eta \omega_r}{2}} \left( \hat{a} - \hat{a}^{\dagger} \right), \qquad
\hat{\phi} = \sqrt{\frac{\hbar}{2 \eta \omega_r}} \left( \hat{a} + \hat{a}^{\dagger} \right), \qquad
\comm{\hat{a}}{\hat{a}^{\dagger}} = 1.
\end{align}
Finally, we obtain the Hamilton operator of the resonator + SNAIL system as 
\begin{align}\label{eq:H-SNAIL-full}
\hat{H}(t) / \hbar = \omega_r \hat{a}^{\dagger} \hat{a} + g_1(t) \left( \hat{a} + \hat{a}^{\dagger} \right) + g_2(t) \left( \hat{a} + \hat{a}^{\dagger} \right)^2
+ g_3(t) \left( \hat{a} + \hat{a}^{\dagger} \right)^3
+ g_4(t) \left( \hat{a} + \hat{a}^{\dagger} \right)^4,
\end{align}
with
\begin{align}
	\label{eq:coeffs_Hamiltonian}
	\hbar g_m(t) = \frac{1}{M^{m-1}} \frac{E_J}{m!} \widetilde{\phi}^{m} 		 	\left\{\begin{array}{lr}
	c_m^{\mathrm{ac}}(t),& \text{for } m=1,2 \\
	c_m^{\mathrm{dc}} + c_m^{\mathrm{ac}}(t),& \text{for } m=3, 4
	\end{array}\right.,
\end{align}
and 
\begin{align}
\widetilde{\phi} = \frac{\cos(\frac{\pi}{2} \frac{\omega_r}{\omega})}{\phi_0} \sqrt{\frac{2 Z_c \hbar}{\pi \frac{\omega_r}{\omega} + \sin(\pi \frac{\omega_r}{\omega})}}
\end{align}
where we used $k d = (\pi / 2) (\omega_r / \omega_0)$ with $\omega_0 = (\pi / 2) (\nu / d)$ and we have made the tunability of couplings through the SNAIL parameters $\alpha$, $n$ and $\Phi_{\rm ext}^{\rm dc}$ and through the modulation $\Phi_{\rm ext}^{\rm ac}(t)$ implicit for readability.

%

\subsection{Undesired Terms and Single Mode Approximation}

From \eqref{eq:H-SNAIL-full} our desired cubic Hamiltonian follows by choosing $\fphi \propto \cos(\omega_r t) + \cos(3 \omega_r t)$ and applying the Rotating Wave Approximation (RWA) in an interaction picture with respect to the resonator - SNAIL free Hamiltonian $\omega_r \hat a^\dagger \hat a$.
As pointed out in the main text, the ideal cubic interaction Hamiltonian,
\begin{align}
\label{eq:cpg-Hamiltonian}
\hat{H}_I &= g_3^{\mathrm{ac}} \left(\hat{a} + \hat{a}^{\dagger} \right)^3,	
\end{align}
is obtained together with terms that are linear and quadratic in the quadrature $\hat{q} \propto \hat{a} + \hat{a}^{\dagger}$.
In the lab frame these terms read
\begin{align}
\label{eq:undesired-terms}
\left( \mathrm{e}^{i \omega_r t} + \mathrm{e}^{-i \omega_r t} + \mathrm{e}^{i 3 \omega_r t} + \mathrm{e}^{- i 3 \omega_r t} \right) \left( \hat{a} + \hat{a}^{\dagger} \right) + \left( \mathrm{e}^{i \omega_r t} + \mathrm{e}^{-i \omega_r t} + \mathrm{e}^{i 3 \omega_r t} + \mathrm{e}^{- i 3 \omega_r t} \right)^2 \left( \hat{a} + \hat{a}^{\dagger} \right)^2,
\end{align}
where we have dropped the explicit coefficients for simplicity.
In the rotating frame and within the RWA
the only resonant terms are
\begin{align}
\label{eq:undesired-terms-rot}
\left(\hat{a} + \hat{a}^{\dagger} \right) + \left[ 3 \left( \hat{a}^2 + \hat{a}^{\dagger 2} + \hat{a} \hat{a}^{\dagger} + \hat{a}^{\dagger} \hat{a} \right) - \left(\hat{a} \hat{a}^{\dagger} + \hat{a}^{\dagger} \hat{a} \right)  \right].
\end{align}
We observe that the terms in the square bracket are equivalent to $3\hat{q}^2/2 - (\hat{a} \hat{a}^{\dagger} + \hat{a}^{\dagger} \hat{a}$).
These residual quadratic and linear terms can be removed or modified by using the Gaussian gates within our universal gate set. This holds e.g. in the case one would like to achieve in particular the cubic gate corresponding to the $\hat T$-gate in GKP encoding~\cite{gottesman_encoding_2001}
\begin{align}
\label{eq:T-gate}
\hat{T} = \mathrm{e}^{\frac{i \pi}{4} \left[ 2 \hat{q}^3/\sqrt{\pi}^3 + \hat{q}^2/\pi - 2 \hat{q}/\sqrt{\pi} \right]}.
\end{align}
In principle, this is possible by flux driving at $\omega_r - \delta$ and $3 (\omega_r - \delta)$ instead of $\omega_r$ and $3 \omega_r$, respectively.
The detuning $\delta$ is then chosen such that it cancels the additional term. \\ 

Also note that one vale of the cubicity is sufficient in order to attain any desired cubicity, provided arbitrary quadratic operations are available. Indeed, the cubicity  can be increased by ``consuming" initial squeezing~\cite{gottesman_encoding_2001, albarelli_resource_2018, takagi_convex_2018}.

In the above analysis we have restricted ourselves to a single resonator mode.
For this approximation to be valid it is necessary %
to avoid populating more than a single resonator mode as all of them are coupled through the SNAIL which may result in additional interaction terms in the Hamiltonian. 
Recalling that, for a standard quarter wavelength resonator, the frequency of the $n$th mode is given by $\omega_n = (\pi \nu / d)(n + 1/2)$, we see that the
frequency relation between the fundamental ($\omega_0$) and the first mode ($\omega_1$) is given by
$\omega_1 = 3\, \omega_0$.
Thus, flux driving at $3 \omega_0$ would in principle lead to populating the mode at $\omega_1$.
However, the presence of the SNAIL modifies this relation as shown above. Furthermore, if this modification is not sufficient one can apply impedance engineering as in \cite{zakka-bajjani_quantum_2011, chang_observation_2020} to obtain non-equidistant mode frequencies.
We therefore conclude that restricting our analysis to a single mode is legitimate.

\subsection{Circuit Parameters}
We now estimate the value of $g_3^{\mathrm{ac}}$ using realistic experimental parameters 
~\cite{frattini_3-wave_2017, frattini_optimizing_2018, sivak_kerr-free_2019, grimm_kerr-cat_2019}.
{\color{black}
These parameters are selected after discussions with experimentalists to prepare a cubic phase state with visible negativity.
As stated in the main text, we expect that the choice of parameters could be improved using classical optimization.
This is however beyond the scope of this work.
}
In particular, we will restrict to a single SNAIL with $n = 3$ large Josephson junctions.
For its design parameters we choose $E_J$ such that $L_J = \SI{600}{\pico\henry}$ and $\alpha = 0.1$.
Choosing $\alpha$ in such a way, results in a vanishing Kerr coupling at $\Phi_{\mathrm{ext}}^{\mathrm{dc}} \approx 0.39 \, \phi_0$. 
To obtain a resonance frequency $\omega_r / 2 \pi$ in the range of \num {4} to \SI{6}{\GHz}, we consider a coplanar waveguide resonator with resonance frequency $\omega_0 / 2 \pi = \SI{8.8}{\GHz}$.
Assuming $Z_c = \SI{50}{\ohm}$ and numerically evaluating Eq.~\eqref{supp-eq:eigenmode-eq} results in $\omega_r / 2 \pi \approx \SI{4}{\GHz}$.
Choosing $\lambda = 1 / 10$ we obtain $\lvert g_3^{\mathrm{ac}} \rvert / 2 \pi \approx \SI{0.28}{\MHz}$ for the coupling that is relevant for the cubic phase gate, while $\lvert g_{3}^{\rm dc} \rvert / 2 \pi \approx \SI{10}{\MHz}$ allowing fast quadratic gates, e.g. squeezing.
{\color{black}
The resulting couplings which are used for the master equation simulation are summarized in Table~\ref{tab:simulation_couplings}.

\begin{table}[!h]
\caption{\label{tab:simulation_couplings} Here $g_m^{\mathrm{dc}}$ denotes the static (dc) couplings of $m$-th order. By $g_m^{\mathrm{ac, 1}}(t) = \tilde{g}_m^{\mathrm{ac, 1}} \varphi_{\mathrm{ext}}^{\mathrm{ac}}(t)$ and $g_m^{\mathrm{ac, 2}}(t) = \tilde{g}_m^{\mathrm{ac, 2}} \varphi_{\mathrm{ext}}^{\mathrm{ac}}(t)^2$ we denote the time-dependent (ac) couplings of $m$-th order that are linear and quadratic in the external modulation, respectively.
Below we give there amplitudes $\tilde{g}_m^{\mathrm{ac, 1}}$ and $\tilde{g}_m^{\mathrm{ac, 2}}$.
As explained in the main text, we assume that $g_1^{\mathrm{ac, 1}}$ can be canceled and therefore have set it to zero.}
\sisetup{round-mode = places,round-precision = 2, zero-decimal-to-integer}%
\begin{ruledtabular}
\begin{tabular}{ccccc}
& $m=1$ & $m=2$ & $m=3$ & $m=4$ \\
$g_m^{\mathrm{dc}} / 2 \pi$ & 	- & 	- & 	\SI{-1.064e+01}{\MHz} & 	\SI{1.396e-01}{\MHz} \\
$\tilde{g}_m^{\mathrm{ac, 1}} / 2 \pi$ & 	\SI{0.0e+00}{\MHz} & 	\SI{-2.669e+01}{\MHz} & 	\SI{5.607e+00}{\MHz} & 	\SI{5.527e-03}{\MHz} \\
$\tilde{g}_m^{\mathrm{ac, 2}} / 2 \pi$ & 	\SI{3.569e+02}{\MHz} & 	\SI{-5.624e+01}{\MHz} & 	\SI{1.165e-02}{\MHz} & 	\SI{-7.391e-02}{\MHz} 
\end{tabular}
\end{ruledtabular}
\end{table}
}
\sisetup{round-mode = off}%

%
\subsection{Master Equation Simulation}

Lastly, we address the impact of the approximations that lead to the Hamiltonian Eq.~(7) of the main text 
taking
Hamiltonian Eq.~\eqref{eq:H-SNAIL-full}
as the starting point.
To this end, we address the preparation of the cubic phase state $\ket{\gamma, r}$ starting from a vacuum state by sequentially applying the squeezing and cubic phase gates.

We begin by describing the preparation of a squeezed vacuum state.
As described in the main text, in this case it is sufficient to apply only a static flux bias $\sphi$ to the SNAIL such that the system Hamiltonian becomes time-independent and reduces to
\begin{align}
\label{eq:SNAIL-H-gaussian}
\hat{H} = \omega_r \hat{a}^{\dagger} \hat{a} + g_3^{\mathrm{dc}} \left(\hat{a} + \hat{a}^{\dagger} \right)^3 + g_4^{\mathrm{dc}} \left(\hat{a} + \hat{a}^{\dagger} \right)^4. 
\end{align}
The squeezing Hamiltonian is obtained by applying an external drive with frequency $\omega_p \approx 2 \omega_r$ to the resonator-SNAIL system~\cite{grimm_kerr-cat_2019}. The driving Hamiltonian with amplitude $\epsilon_d$ has the form $\hat{H}_d = \epsilon_d \sin(\omega_p t) (\hat{a} + \hat{a}^{\dagger})$ up to a phase that determines the squeezed quadrature. Here we choose the phase such that we obtain squeezing in the $\hat{p}$ quadrature in accordance with the definition of the ideal cubic phase state~\cite{gottesman_encoding_2001}.
The drive will lead to an effective displacement of the resonator field enabling three-wave mixing. 
In the regime where $\lvert \omega_p \pm \omega_r \rvert \gg \kappa$, with $\kappa$ the single photon loss rate, the effective displacement amplitude can be written as~\cite{grimm_kerr-cat_2019}
\begin{align}
\label{eq:eff-displacement}
\xi_{\rm eff}(t) \approx i \frac{1}{3} \frac{\epsilon_d}{\omega_r} \mathrm{e}^{i 2 \omega_r t} \equiv i \overline{\xi}_{\rm eff} \mathrm{e}^{-i 2 \omega_r t},
\end{align}
where we used $\omega_p \approx 2 \omega_r$. In our simulation we choose $\epsilon_d$ such that $\overline{\xi}_{\rm eff} \approx -0.125$.
To obtain the squeezed state we do a master equation simulation using QuTiP~\cite{johansson_qutip_2013} with Hamiltonian parameters as discussed above, loss rate $\kappa / 2 \pi = \SI{50}{\kHz}$ and initializing the resonator in the vacuum state. The evolution time $t_{\rm sq} \approx \SI{14}{\ns}$ is chosen to obtain a squeezing parameter $r \approx 0.7\, (\SI{6}{\dB})$. We quantify the amount of squeezing by calculating $r =- \log(V / V_0) / 2$ with $V = \langle \hat{p}^2 \rangle - \langle \hat{p} \rangle^2$ the variance of the $\hat{p}$ quadrature and $V_0$ the variance of the vacuum state.

After preparing the system into a squeezed state we apply the cubic phase gate by modulating the external flux.
In the presence of the modulation, our system is described by
the Hamiltonian Eq.~\eqref{eq:H-SNAIL-full}, where we include into the coefficients the contributions from $c_n^{\mathrm{dc}}$ (only for $n=3, 4$) and the contributions from $c_n^{\mathrm{ac}}$ up to second order in $\lambda$, i.e. the small modulation amplitude (see also Eq.~\eqref{eq:coeffs_Hamiltonian}).
Here, we consider a modulation of the form $\fphi/\phi_0 = \lambda [ \cos(\widetilde{\omega}_r t) + \cos(3 \widetilde{\omega}_r t)]$ with $\widetilde{\omega}_r = \omega_r - \delta \omega$ in order to take into account the small but finite shift of the resonance frequency due to the term $\propto g_2(t)$.
We use $\delta \omega = -2 g_2^{\mathrm{ac,2}} \approx 2 \pi \times \SI{1.1}{\MHz}$, where by $g_2^{\mathrm{ac,2}}(t)$ we refer to the coupling to $( \hat{a} + \hat{a}^{\dagger})^2$ that is quadratic in the modulation $\fphi$ and thus selects a frequency shift resonantly (compare also Eqs.~\eqref{eq:undesired-terms} and \eqref{eq:undesired-terms-rot}).
For the term $\propto g_1(t)$ we neglect the contribution linear in $\lambda$ since its effect can be corrected via a displacement of the resonator field by an external current source.

To obtain a cubicity of $\gamma \approx 0.1$ 
we evolve the open system 
for a time $t_g \approx \SI{19}{\ns}$.
Before comparing the obtained state to an ideal cubic phase state $\ket{\gamma, r}$ with matched cubicity $\gamma$ and squeezing $r$ we apply a unitary rotation and displacement operation.
The role of the former is to reverse the rotation due to the evolution in the lab frame.
The latter reverses the effective displacement created during the squeezed state preparation given by $\overline{\xi}_{\rm eff}$.

\begin{figure}[!htp]
    \centering
    \includegraphics{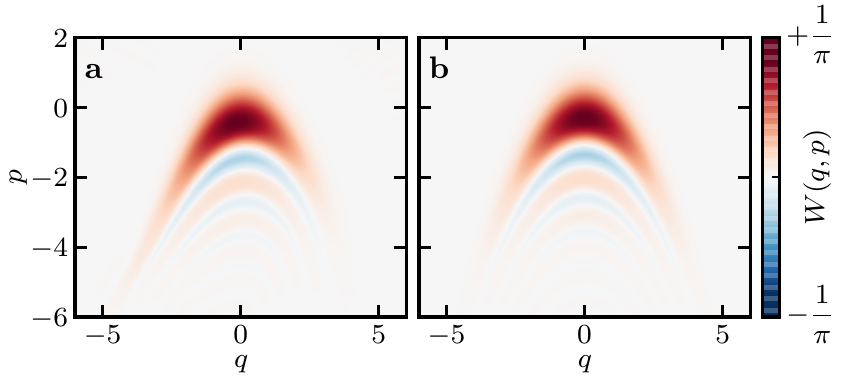}
	\caption{Wigner distribution for the cubic phase state $\ket{\gamma, r}$ (This is the same as Figure~2 of the main text). \textbf{a}~Obtained from a master equation simulation with the Hamiltonian \eqref{eq:H-SNAIL-full} and single photon loss rate $\kappa / 2 \pi= \SI{50}{\kHz}$ by sequentially applying the squeezing and cubic phase gate to an initial vacuum state $\ket{0}$. \textbf{b}~Ideal cubic phase state with matched cubicity $\gamma \approx 0.1$ and squeezing $r \approx 0.70 \; (\approx \SI{6}{\dB})$.}
    \label{fig:cubic_phase_state_full}
\end{figure}

In Figure~\ref{fig:cubic_phase_state_full} we show the Wigner distributions of the final state and of the ideal cubic phase state with matched parameters.
We observe that the created state shares many of the significant properties of the ideal cubic phase state.
A qualitative difference is that the small undesired terms in the Hamiltonian Eq.~\eqref{eq:H-SNAIL-full} lead to a small tilt or smearing out to the left (negative $q$ values) breaking the symmetry along $q=0$.
Nevertheless, the obtained state contains most features of the ideal cubic phase state which is verified by calculating the fidelity with respect to the ideal state in Figure~\ref{fig:cubic_phase_state_full}b yielding \SI{97.2}{\percent}.

\begin{figure}[!htp]
    \centering
    \includegraphics{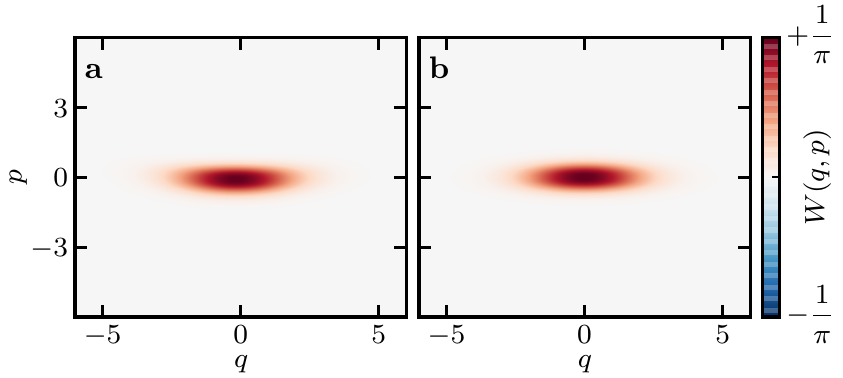}
	\caption{Wigner distribution for the squeezed state $\ket{r}$. \textbf{a}~Obtained from a master equation simulation with the Hamiltonian \eqref{eq:SNAIL-H-gaussian} and single photon loss rate $\kappa / 2 \pi= \SI{50}{\kHz}$ from an initial vacuum state $\ket{0}$. \textbf{b}~Ideal squeezed state with matched squeezing $r \approx 0.70 \; (\approx \SI{6}{\dB})$.}
    \label{fig:squeezed_state}
\end{figure}

For completeness, in Figure~\ref{fig:squeezed_state}  we show  the Wigner distributions of the corresponding (intermediate) squeezed states.
The state obtained from evolution under Hamiltonian Eq.~\eqref{eq:SNAIL-H-gaussian} in Figure~\ref{fig:squeezed_state}a is for this purpose transformed to the rotated-displaced frame as explained above.
The fidelity to the ideal state is \SI{99.1}{\percent}.

%
